\documentclass[amsmath, amssymb, twocolumn, pra, aps]{revtex4-2}

\usepackage{newtxtext, newtxmath}
\usepackage{graphicx}
\usepackage{dcolumn}
\usepackage{svg}
\usepackage{bm}
\usepackage{wrapfig}
\usepackage{xcolor}
\usepackage{placeins}
\usepackage{hyperref}
\hypersetup{colorlinks, citecolor=blue,
   		 	filecolor=black, linkcolor=blue,
    		urlcolor=blue}
\usepackage{array}
\usepackage[capitalise]{cleveref}

\begin{document}

\title{Collision Parameters Needed to Measure Polarisation-Dependent Nonlinear Pair Creation}

\author{Mathias Samuelsson}
\email{mathias.samuelsson@physics.gu.se}
\author{Thomas G. Blackburn}
\email{tom.blackburn@physics.gu.se}
\affiliation{Department of Physics, University of Gothenburg, SE-41296 Gothenburg, Sweden}

\date{\today}

\begin{abstract}
Pair creation by a $\gamma$ ray in a high-intensity electromagnetic field (the nonlinear Breit-Wheeler process) is sensitive to the $\gamma$-ray polarisation. Here we study the stability required in order to measure this polarisation dependence in a head-on collision between a high-power linearly polarised laser and a relativistic electron beam, where the $\gamma$-rays produced via nonlinear Compton scattering are highly polarised. We find that the laser strength parameter $a_0$ has to be known to within a few per cent, and that the alignment of the laser and electron beam cannot fluctuate by more than half the spot size, in order to attribute the reduction in positron yield to polarisation effects. These collision parameters are difficult to achieve, but may be realised in future precision experiments that focus on beam stability.
\end{abstract}

\maketitle
\section{\label{sec:intro}Introduction \text{ }}
Current high-power lasers can now reach intensities of $10^{23}$ W/cm$^2$~\cite{Yoon:21}, which enables us to conduct experiments investigating extreme electromagnetic conditions~\cite{piazza,chargedpart,FEDOTOV20231}.
In such intense laser fields, the interaction of charges with the background field becomes nonperturbative and must be treated exactly \cite{Wolkow1935berEK}. The strength of this interaction is characterised by the parameter $a_0 = \frac{eE_0}{mc\omega}$ (the normalised amplitude of the vector potential) where $e$ is the elementary charge, $E_0 = |\mathbf{E}| = c|\mathbf{B}|$ the electric field, $m$ is the electron mass and $c$ is the speed of light. The $a_0$ parameter characterises how nonlinear (multiphoton) the interaction is, because it tells us how many laser photons interact with the charge \cite{Nikishov:1964zza}. Similarly, the additional parameter that characterises the importance of quantum effects is the quantum nonlinearity parameter,

\begin{align}
    \chi &= \frac{\hbar}{E_S mc} \sqrt{|F_{\mu \nu}p^{\nu}|^2} \label{eq:chi}.
\end{align}
Here, $E_S\approx 1.3\cdot10 ^{18}$ V/m is the critical or Schwinger field where QED turns nonlinear~\cite{PhysRev.82.664} (although previously known by \citet{Sauter1931}), $F_{\mu \nu}$ is the field tensor and $p^\nu$ is the electron four-momentum. 

In laser fields where $a_0 \gg 1$ and $\chi \sim 1$ we see nonlinear Compton scattering, where an electron absorbs many laser photons and emits a high-energy $\gamma$ photon, and nonlinear pair creation (also known as the Breit-Wheeler process) where a high-energy $\gamma$ photon interacts with the laser photons and creates an electron-positron pair \cite{breit,reiss}.
These events can occur in extreme astrophysical environments, such as accretion discs of black holes and magnetospheres of pulsars \cite{Hibschman_2001,Harding_2006,Timokhin_2019,Olausen_2014}.
In laboratories, the conditions required to reach the nonlinear regime of QED are possible in laser-electron colliders, where a high-power laser and a high-energy electron beam counter propagate. The E144 experiment was the first to observe moderately nonlinear Breit-Wheeler using an electron beam of energy 46.6 GeV and laser intensities up to $10^{18}$ W/cm$^2$ ($a_0 \simeq 0.4$) \cite{bula1996,lightbylight_burke}.
With increasing intensities since the petawatt revolution, made possible with chirped pulse amplification~\cite{STRICKLAND1985219,lasermap}, we can now go further into the nonlinear regime. Upcoming facilities~\cite{appollon,zesus,vulcan,ELIBP, ELINP} aim to measure pair creation in the strongly nonlinear regime, $a_0 \gg 1$, which is yet to be accomplished~\cite{Blackburn_2018,Hartin_19,yan-fei_2020,Mercuri-Baron_2021,Salgado_2021,Eckey_2022,ELI_2023,Martinez_2023,Pouyez_2024,Barbosa_2024}.

Part of doing precision tests of QED is measuring how the rates depend on the parameters $a_0$ and $\chi$. However, there is another parameter that matters to pair creation, namely the $\gamma$-ray polarisation. It has been shown this changes the electron-positron yield in a laser-electron collision if the laser is linearly polarised \cite{bking,Feng_2020,ya-nan_2021,Qian_2023,Tang_2023}.
Here, we investigate the collision parameters needed to identify the polarisation dependence of nonlinear pair creation with confidence. We perform simulations of laser-electron beam collisions, where $\gamma$-ray polarisation is resolved, for the capabilities of current and planned laser facilities. The results show that the positron yield is reduced by 12\% due to $\gamma$-ray polarisation, approximately independently of $a_0$ and electron energy. In order to measure this difference we find that $a_0$ must be stable to within few per cent and that the alignment of the laser and electron beams cannot fluctuate by more than half the focal spot size.

\section{\label{seq:rates}Polarised photons and pair creation}
\begin{figure}
    \includegraphics[width=0.8\linewidth]{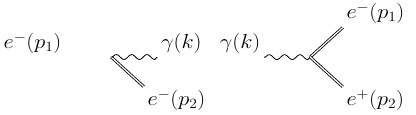}
	\caption{The nonlinear Compton scattering (left) and nonlinear Breit-Wheeler process (right) where a high-energy photon is emitted from Compton scattering and interact with the laser photons and produce a electron-positron pair. Double fermion lines indicate that multiple photons interact due to the strong laser field.}
    \label{fig:feynmandiagram}
\end{figure}

In strong background fields, high-energy photons create electron-positron pairs. These newly created pairs are still in a strong accelerating field, meaning that they may emit high-energy photons and these photons can pair create as well. This process, where a chain of pair creation and photon emission occurs, is called a ``shower'' when the supplied energy comes from the initial particles. If the strong background field, e.g. the laser, supplies the particles with energy it can become an ``avalanche'' \cite{MIRONOV20143254}.
To further understand these phenomena we consider the two subprocesses for electromagnetic showers which are nonlinear Compton scattering and nonlinear Breit-Wheeler (pair creation). In nonlinear Compton scattering, the background photons are absorbed by an electron, which emits a high energy photon. This happens in the following scheme: $e^-(p_1) + n\omega_0 \to e^-(p_2) + \gamma(k)$, where $k$ is the (high-energy) photon momentum, $p_{1,2}$ are the momenta of the electron before and after, and $n$ is the number of absorbed laser photons with energy $\omega_0$. This is shown in \cref{fig:feynmandiagram}(left) where double fermion lines represent the interaction with many laser photons. In nonlinear Breit-Wheeler, a high-energy photon interacts with the background to create an electron-positron pair $\gamma(k) + n\omega_0 \to e^-(p_1) + e^+(p_2)$, accordingly (see right-hand side of \cref{fig:feynmandiagram}).

If the laser is linearly polarised there is a asymmetry in the transverse directions as the laser electric field oscillates in a single plane. Therefore it matters to the two subprocesses whether the high-energy photon is polarised along the electric field (which we call E-polarised) or along the magnetic field (B-polarised). In contrast, in a circularly polarised laser the electric field rotates in a circle, which restores the symmetry in the transverse directions.

Nonlinear Compton scattering has different properties depending on the laser polarisation. For a circularly polarised laser the emitted high-energy photons will take on the circular polarisation of the laser field (if $a_0$ is not very large), and for a linearly polarised laser they will be linearly polarised. We can understand this classically from the electron motion, which must be parallel to the field oscillation. Therefore we expect the high-energy photons, which come from back scattering, to be mostly E-polarised.

This becomes important as nonlinear Breit-Wheeler depends on the photon polarisation, especially for linearly polarised lasers \cite{Ritus1985}. To understand this we may study the rate at which pair creation occurs, which is given by \cref{eq:paircreat} in the locally constant field approximation (LCFA):

\begin{align}
    \frac{\mathrm{d}^2 W^{ \pm}}{\mathrm{d} f \mathrm{~d} \zeta}=\frac{\alpha \delta}{\sqrt{3} \pi \tau_C} \left[1+\zeta^{2 / 3}\left(\frac{f}{1-f}+\frac{1-f}{f}-S_1\right)\right] K_{1/3}(\delta \zeta). \label{eq:paircreat}
\end{align}
Here $\alpha$ is the fine structure constant, $\tau_C$ is the Compton time, and $\delta = \frac{2}{3\chi_\gamma f(1-f)}$, where  $f= \hbar\omega/\mathcal{E}$ is the ratio of the photon energy ($\omega$) and electron energy ($\mathcal{E}$), $\zeta \approx (1 + \theta^2\gamma ^2)^{3/2}$ is a transformed polar angle, and $S_1$ is a Stokes parameter of the photon \cite{10.1063/5.0159963}. It takes the values $S_1 = -1$, meaning that the photon polarisation is perpendicular to $\mathbf{E}_\perp + \mathbf{v}\times \mathbf{B}$ (the effective acceleration, where $\mathbf{E}_\perp$ is the component that is perpendicular to the velocity $\mathbf{v}$), $S_1 = 1$ which means that the photon polarisation is parallel to the acceleration and $S_1 = 0$ meaning it is unpolarised. From this we see that B-polarised photons are more likely to pair create than E-polarised photons. However, B-polarised photons are less likely to be emitted from nonlinear Compton scattering.

It has been common practice to take the emitted photons from nonlinear Compton scattering to be unpolarised, but as \citet{bking} discuss, and what we see from the rate of pair creation, this cannot be a valid assumption when using a linearly polarised laser as the emitted photon \textit{is} polarised and pair creation \textit{depends} on the photons polarisation. We know that taking the photon polarisation into account changes the pair yield because the two subprocesses depend on the photon polarisation \cite{bking,Feng_2020,Qian_2023,ya-nan_2021,Tang_2023}.
The purpose of this paper is to determine if this polarisation-induced difference is measurable in an experiment which collides a laser with an electron beam to produce a electromagnetic shower.

\section{\label{sec:results}Results }

\subsection*{Simulation setup}
We simulate 1D head-on collisions between an electron beam and a linearly polarised laser using the Ptarmigan simulation code \cite{10.1063/5.0159963}. The electron beam consists of $N\sim10 ^{5}-10^{6}$ primary electrons with a weight of $N^{-1}$ (so that the yield is normalised per electron). The electrons have an energy in the range of $1.0 < \mathcal{E}< 22.0$ GeV and are in a dense Gaussian package of length and radius 1 $\mu$m, respectively. These electrons collide head-on with an infrared, linearly polarised laser with a wavelength of 0.8 $\mu$m and a Gaussian intensity profile with a FWHM duration of 30 fs. The normalised amplitude ranges between $5.0 < a_0 < 100.0$ to cover present and future lasers. In the control settings for Ptarmigan we specify LCFA because for this range of $a_0$ LCFA should be sufficiently accurate. Pair creation and radiation reaction is naturally included, though since the pair-creation probability is very small in the low $\chi$ regime, we increase the rate by a large number, while reducing the macroparticle weight by the same factor, to resolve the positron yield~\cite{Blackburn_2022}. In order to study polarisation effects, we require that we can distinguish the polarisation of the photons from Compton scattering, which Ptarmigan does, and that it can be configured to take photon polarisation into account in the pair-creation rate.

\subsection{\label{sec:gammas} Photon emission}

\begin{figure*}
    \centering
    \includegraphics[width=0.8\linewidth]{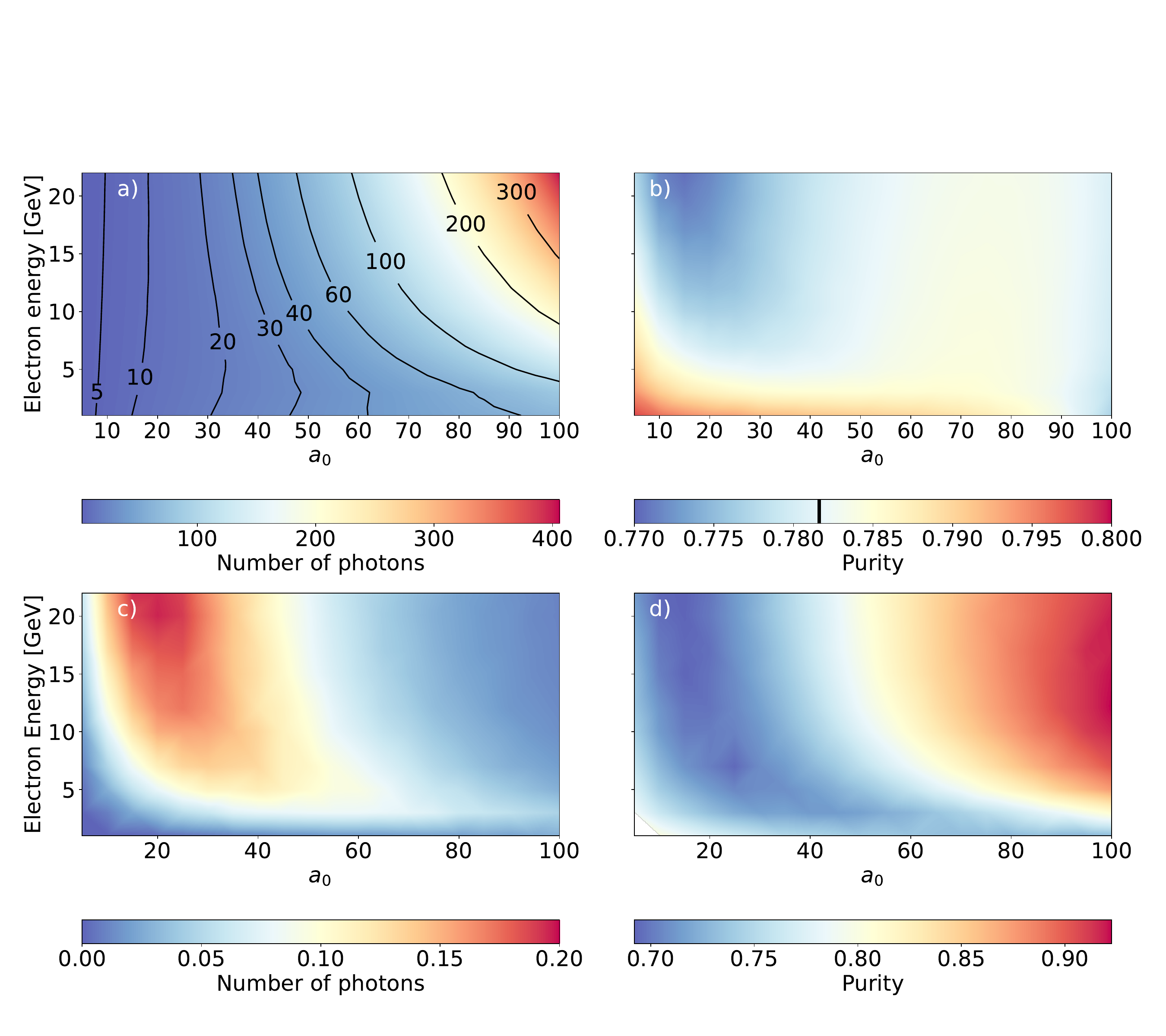}
    \caption{a) The total number of all photons (per electron) as a function of electron energy $\mathcal{E}$ and laser amplitude $a_0$. b) The E-polarisation degree (purity) of all photons. c) The total number of energy photons (per electron) with energy $>\mathcal{E}/2$ and d) the E-polarization degree of these photons.}
    \label{fig:Nphotons}
\end{figure*}

The total number of photons produced from the electron beam and radiating electron-positron pairs is shown in \cref{fig:Nphotons}a) over different electron energies $\mathcal{E}$ and laser amplitudes $a_0$. In this plot we see that for very low $a_0$ that the contours are nearly vertical which indicates a weak dependence on $\mathcal{E}$. The photon emission rate is
\begin{align}
    \frac{d N_\gamma}{dt} = \label{eq:Ngamma_prop} \frac{\alpha}{\tau_C}\frac{1}{\gamma}
    \begin{cases}
        \frac{5}{2\sqrt{3}}\chi,\quad &\chi \ll 1\\
        \frac{14\Gamma(\frac{2}{3})}{9\sqrt[3]{3}}\chi^{\frac{2}{3}}, \quad &\chi\gg1, 
    \end{cases}
\end{align}
where $\gamma$ is the electron Lorentz factor and $\Gamma(\frac{2}{3}) \simeq 1.35$~\cite{Ritus1985}.
If $\chi$ is very small the rate scales linearly on $a_0$ and there is no energy dependence.

As $a_0$ increases up to $50$ the numbers of photons increase substantially and now the contours show a nonlinear dependence on $\mathcal{E}$ and $a_0$, due to quantum corrections as seen in \cref{eq:Ngamma_prop}. If $a_0$ increases even further we enter the cascade regime, where there is an exponential growth of photons due prolific pair creation (which will be discussed in the next section).

In \cref{fig:Nphotons}b) we show the degree of E-polarisation of the photons, which is defined,
\begin{align}
p_{E} =  \frac{N_{\gamma,E}}{N_{\gamma,E} + N_{\gamma,B}} = \frac{1 + \langle S_1 \rangle}{2},   \label{eq:purity}
\end{align}
where $N_{\gamma,E/B}$ is the number of photons that are either $E$-, or $B$-polarised respectively and $\langle S_1 \rangle$ is the averaged Stokes parameter. We see that the degree of E-polarisation is effectively constant at 0.78 over the entire space. However, at low $\chi$ we see that the purity starts to increase because quantum effects, such as spin flips and recoil, become less important. Crucially, the degree of E-polarisation is much larger than 0.5, which corresponds to unpolarised photons, and this will affect the pair yield.

In \cref{fig:Nphotons}c) the number of photons with more than half of the initial electron energy is shown. The number of high energy photons, in contrast to \cref{fig:Nphotons}a), is significantly lower because the spectrum is suppressed at high energies. With increasing $a_0$ the number of high-energy photons first increases, because nonlinear effects become more important, and then decreases, because radiation reaction effects lower the electron energy and pair creation consumes the photons.

The purity of high energy photons is shown in \cref{fig:Nphotons}d). In the low $\chi$ regime, the purity initially decreases with increasing $\chi$, because of quantum corrections, exactly as in \cref{fig:Nphotons}b). As $\chi$ increases, the purity starts to increase again. This may be explained by vacuum dichroism: B-polarised photons pair create at a faster rate then E-polarised \cref{eq:paircreat}, so in the region $a_0>50$, most of the B-polarised photons will have decayed, leaving the E-polarized photons behind, which increases the purity \cite{PhysRev.136.B1540, PhysRevLett.119.250403, PhysRevD.111.036025}. \Cref{fig:Nphotons}b) also shows a nonuniformity for large $a_0$, due to combination of radiation reaction, which lowers $\chi$, and vacuum dichroism.

\subsection{\label{sec:positrons} Pair creation}

\begin{figure*}[!ht]
    \centering
    \includegraphics[width=0.8\linewidth]{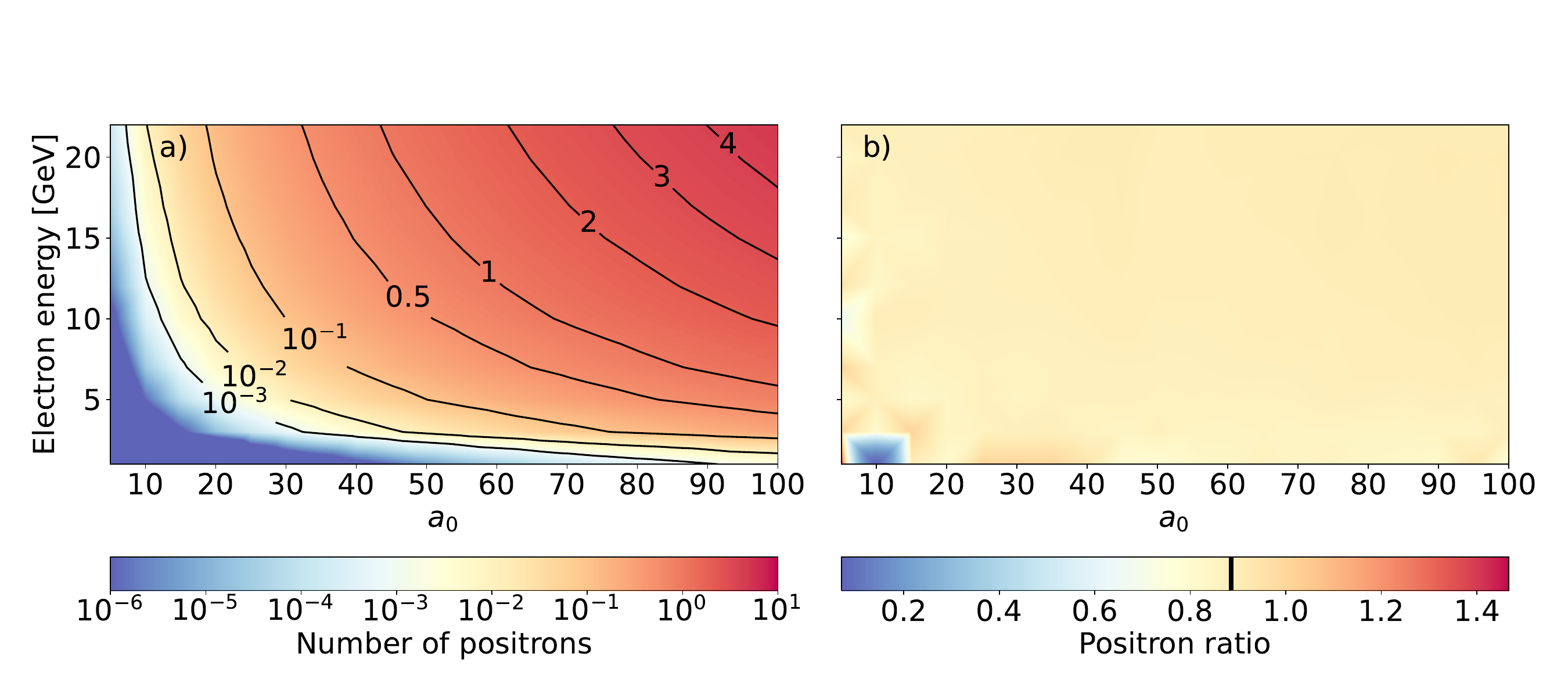}
    \caption{a) The total number of positrons (per electron) as a function of electron energy $\mathcal{E}$ and laser amplitude $a_0$ (log-scaled).
    b) The ratio of the positron yield when photon polarisation is taken into account and the positron yield when photons are assumed to be unpolarised. The black line in the colour bar indicates the average value of 0.88.}
    \label{fig:Npositrons}
\end{figure*}

Both \cref{fig:Nphotons}a-b) tell us that the assumption that the hard photons which create pairs are unpolarised is not valid, and to see how the polarisation affects pair creation we study the positrons.

The total number of positrons is shown in \cref{fig:Npositrons} as a function of $\mathcal{E}$ and $a_0$. We see that increases by orders of magnitude, from ${<}10^{-6}$ to $5$ positrons per electron because the rate is exponential in $\chi$. This explains why the number of photons increases so dramatically in \cref{fig:Nphotons}a). When the Ptarmigan simulates the interaction, it takes the polarisation of the photons into account when calculating the pair creation rate. If the inclusion of photon polarisation is ignored then we get the positron yield as if the photons were unpolarised. We compare these two cases in \cref{fig:Npositrons}b), where we show the ratio of the two positron yields. We see that the ratio is effectively constant at $\approx0.886$, showing that in the polarised case there are 12\% fewer positrons. (Along the axes there are some artifacts due to the very small positron yield, which makes it difficult to resolve the ratio.) 

Let us show that this positron ratio is consistent with analytical calculations. The total positron production rate may be written as a linear combination of $W_E$ and $W_B$ scaled by their respective purities as defined in \cref{eq:purity} i.e. $W^\pm  = p_E W_E + p_B W_B$. Here $W_E$ and $W_B$ are the positron rates for purely E-polarised photons and B-polarised photons, respectively. The ratio is calculated with respect to the positron rate for unpolarised photons, $W^\pm_\text{Unpol} = \frac{1}{2}\left(W_E +W_B\right)$. By integrating \cref{eq:paircreat}, we find that for $\chi \ll 1$ the ratio $W_B/W_E =2$ and for $\chi \gg 1$ it is $W_B/W_E = \frac{3}{2}$ \cite[Sec. 3.2]{baier.1998}.
From this we compute the positron ratios given the asymptotic bounds of $W_B/W_E$ and the purity found in \cref{sec:gammas}, which is $p_E \approx 0.78$. In the limit of $\chi \ll1$ the ratio is close to 0.815 and for $\chi \gg 1$ the ratio is close to 0.885. These lower and upper bounds correspond well with the simulated average positron ratio of 0.886. Here we get a higher value than analytical calculations suggest, most likely because our naive approach to calculating the purity includes photons of all energies while only high-energy photons are likely to pair create.

The results of \cref{fig:Nphotons}b) and \cref{fig:Npositrons}b) clearly state that the hard photons \textit{are} polarised and most of them (78\%) are E-polarised. Unpolarised hard photons produce 12\% more positrons than polarised photons. This difference in yield should be big enough to be noticeable, but now the question is what is required of an experiment in order to identify this polarisation-induced  reduction in a laboratory.

\subsection{\label{sec:laserint}Electron energy and laser intensity tolerance}
We have seen that treating photons as unpolarised causes an overestimation of the number of positrons, on average by 12\%. However, a reduction may be caused by a lower electron beam energy, a lower $a_0$ or a misalignment in space and time. 

\begin{figure}
    \centering
    \includegraphics[width=1\linewidth]{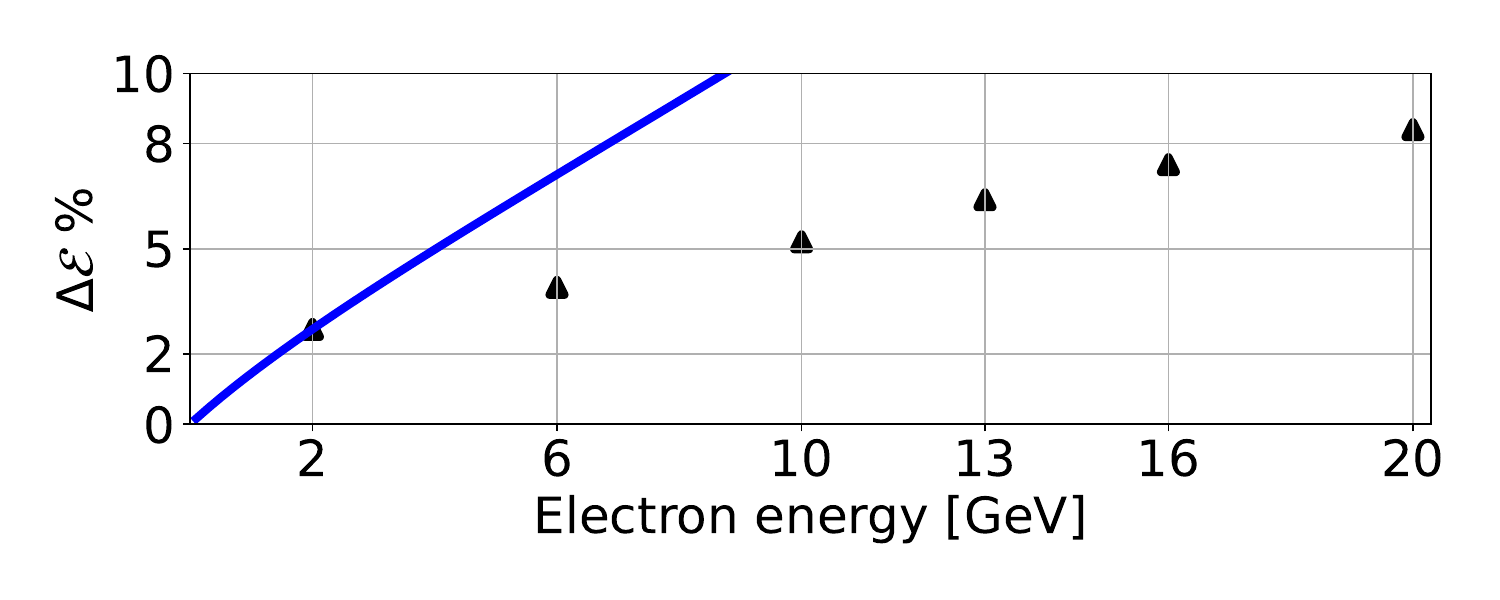}
    \caption{The maximum permitted fluctuation in $\mathcal{E}$ that allows the reduction in the yield due to photon-polarisation dependence to be visible, from simulations (dots) and theory (line).
    The amplitude of the counter-propagating laser is fixed to be $a_0=30$.}
    \label{fig:devE}
\end{figure}

Let us consider how the positron yield changes due to fluctuations in electron beam energy $\mathcal{E}$. By simulating the interaction for a small range of energies around a nominal energy, we may find the fluctuation that causes a reduction in the positron yield of 12\%, which would obscure the polarization signal. The results of 1D simulations are shown in \cref{fig:devE} as a maximum energy fluctuation ($\Delta\mathcal{E}$) in per cent over a range of electron energies at a fixed $a_0 = 30$. We see that allowed fluctuation increases with electron energy but only to a few per cent.

This can be analytically verified by considering the probability to create a pair from a photon with energy $\omega$, which we take to be the equal to the electron energy $\mathcal{E}$,
\begin{align}
    P_+(\omega) &= \alpha n a_0 R(\chi)\label{eq:total_prob}.
\end{align}
where $\chi = 2\omega \omega_0 a_0 \hbar^2/(m^2 c^4)$ is the quantum nonlinearity parameter, $R(\chi)$ is an auxiliary function given in \cite{PhysRevA.96.022128}, and $n = c\tau/\lambda$ is the number of wavelengths in a FWHM pulse duration $\tau$. From this we get the fluctuation of $\mathcal{E}$ for a given $\delta P$ (the absolute change in probability), 
\begin{align}
    \frac{\delta\mathcal{E}}{\mathcal{E}} = \frac{\delta P}{P}\left(\frac{\chi}{R(\chi)}\frac{\partial R(\chi)}{\partial\chi}\right)^{-1}\label{eq:devE},
\end{align}
The prediction of \cref{eq:devE} are shown in \cref{fig:devE} as the solid line. For very low energies the agreement with simulation is good, but it deviates quickly from the simulated result. This may be explained by the fact that \cref{eq:total_prob} only includes pair creation from an \textit{existing} photon beam i.e. it does not include Compton scattering in the process, which is part of the simulation. Nonlinear Compton scattering creates photons with a broad range of energies. At low electron energies only the highest energy photons ($\omega\simeq\mathcal{E}$) may pair create, but as the electron energy increases, more of the lower energy photons will contribute.

These restrictions on the electron energy fluctuations are rather strict and may not be achievable in present day laser wakefield accelerators at such energies, and therefore we consider the use of conventional linear accelerators, which have near perfect energy stability and very low energy spread \cite{gonsalves,Abramowicz2021}.

\begin{figure}
    \centering
    \includegraphics[width = 1\linewidth]{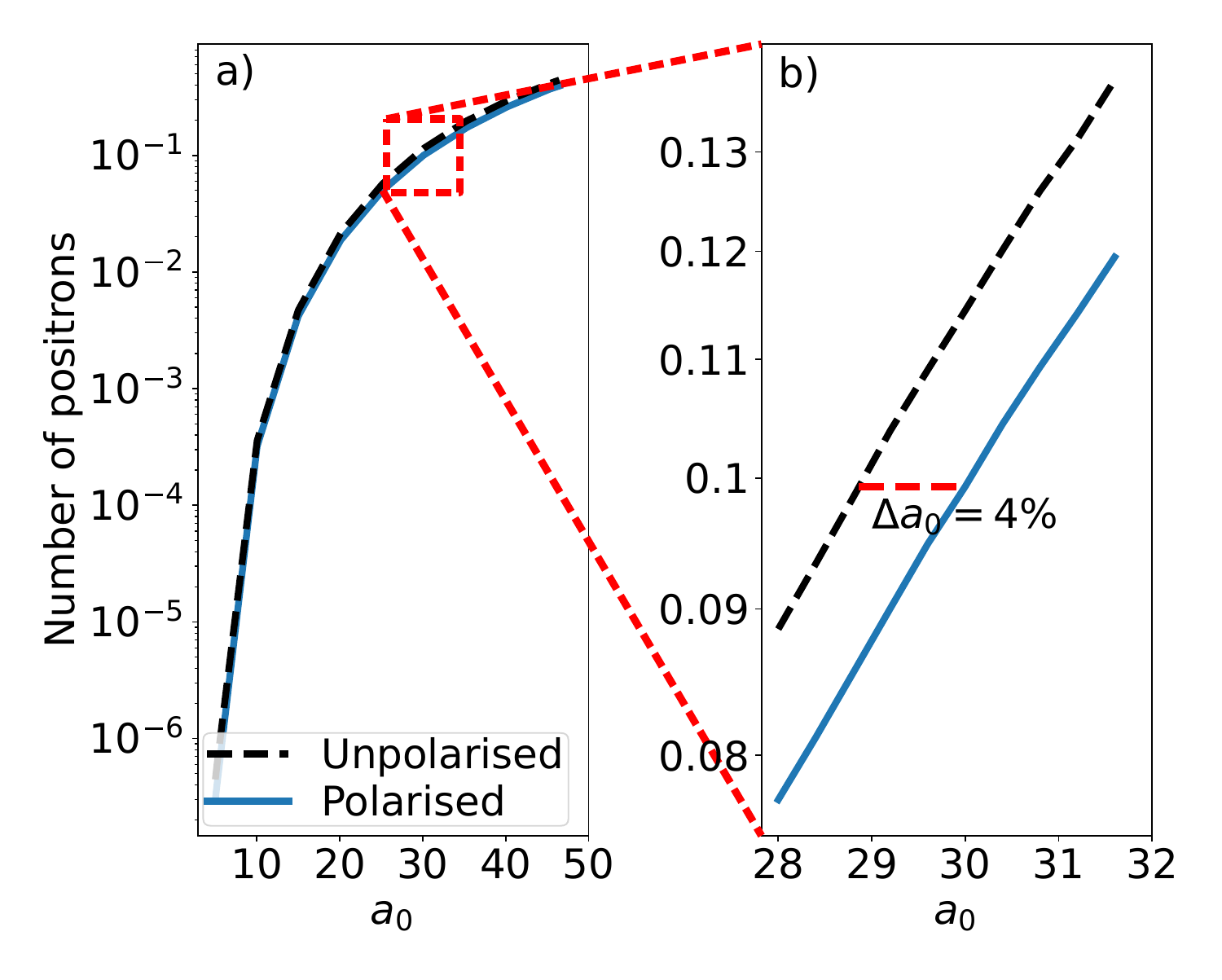}
    \caption{The number of positrons as a function of $a_0$ for a 10-GeV electron beam, a) over the complete range of $a_0$ and b) inside the region marked with a red-dashed rectangle. The difference in positron yield is 12\% and the equivalent reduction in $a_0$ is 4\%.}
    \label{fig:polarisation_a0_measure}
\end{figure}

A reduction in the positron yield can also be caused by a lower $a_0$, which may itself be caused by a fluctuation in the laser energy, spatial or spectral phase, or a misalignment in space and time. We now consider how the positron yield changes with respect to a general fluctuation in $a_0$. As an example, let us consider the number of positrons produced by a 10 GeV electron beam and a laser with $a_0 \sim 30$, as shown in \cref{fig:polarisation_a0_measure}. We can see that it is possible to get the same change in positron yield between the unpolarised and polarised cases (0.114 and 0.099 positrons respectively) by lowering $a_0$ by 4\%. In order to see that this difference is caused by photon polarisation, we require any fluctuation in $a_0$ to be smaller than 4\%.

\begin{figure}
    \centering
    \includegraphics[width = 1\linewidth]{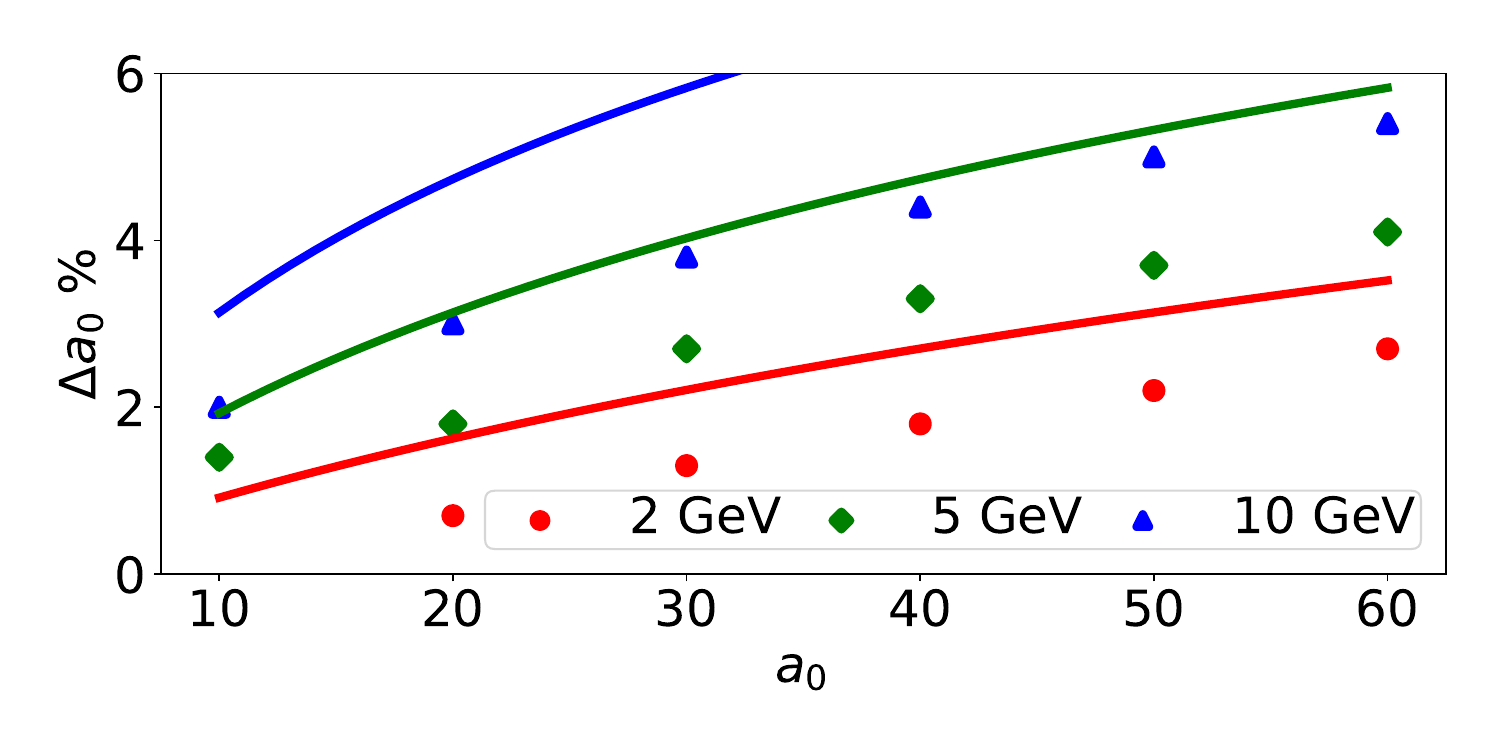}
    \caption{The maximum permitted fluctuation in $a_0$ that allows the reduction in the yield due to photon-polarisation dependence to be visible, from simulations (points) and theory (lines).}
    \label{fig:deviate_a0}
\end{figure}

The largest allowed fluctuation in $a_0$ such that polarisation effects can be observed is extended to a wider range of $a_0$ and $\mathcal{E}$ in \cref{fig:deviate_a0}. It grows with both $a_0$ and electron energy because the positron yield grows more slowly at larger $\chi$ (as seen in \cref{fig:polarisation_a0_measure}). Although the largest allowed fluctuations increase, $a_0$ still has to be known up to a few per cent.

From \cref{eq:total_prob} we get the fluctuation of $a_0$ for a given $\delta P$ (the absolute change in probability) in the same way as \cref{eq:devE},
\begin{align}
      \frac{\delta a_0}{a_0} &= \frac{\delta P}{P}\left(1 +\frac{\chi}{R(\chi)} \frac{\partial R(\chi)}{\partial \chi}\right) ^{-1}\label{eq:a0fluct}.
\end{align}

The predictions of \cref{eq:a0fluct} are shown in \cref{fig:deviate_a0} as solid lines. The predictions are bigger than the simulation results, but in general follow the same trend. The scaling is different because we have assumed that the photons have the same energy as the electrons, whereas in simulation we start from an electron beam which undergoes Compton scattering.

We note that \cref{eq:devE} and \cref{eq:a0fluct} are very similar, but the latter contains an extra term in the denominator, meaning that the allowed fluctuation is smaller. This implies that, while electron energy may fluctuate too much for polarization dependence to be seen, controlling the fluctuation in $a_0$ is more important as it is more strongly bound. For example, at 10 GeV and $a_0=30$ the allowed fluctuations from simulation are $\Delta\mathcal{E}\approx 5.2\%$ and $\Delta a_0 \approx 3.8 \%$, whereas theory predicts $\Delta \mathcal{E}\approx 11.3\%$ and $\Delta a_0 \approx 5.8\%$. While the magnitude of the fluctuations does not agree, the theory is consistent in that $\Delta\mathcal{E}>\Delta a_0$.

\begin{figure}
    \centering
    \includegraphics[width= 1\linewidth]{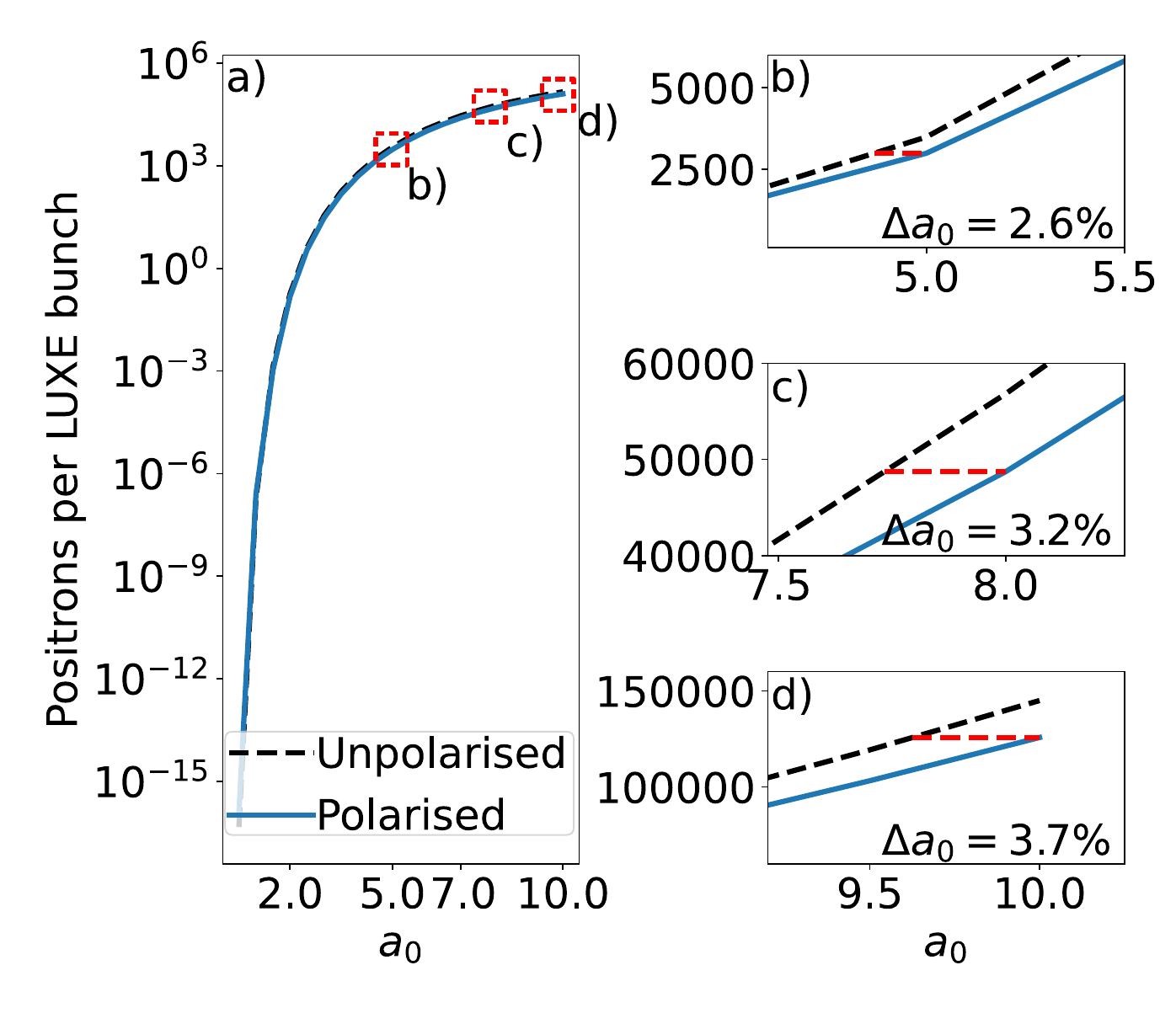}
    \caption{The positron yield per bunch crossing expected for the LUXE experiment, for a) polarised (solid) and unpolarised (black, dashed) photons. b-d) Specific regions from a) to show the maximal fluctuation of $a_0$ that is allowed.}
    \label{fig:LUXEsweep}
\end{figure}

Finally, let us apply this analysis to the LUXE experiment, which intends to collide a powerful laser with the European XFEL electron beam in order to study nonlinear QED \cite{Abramowicz2021}. We simulate the LUXE setup where the electron beam has a radius of 5 $\mu$m, energy of $\mathcal{E}_0= 16.5$ GeV and the number of electrons is $1.5 \cdot 10^{9}$. The laser amplitude is $a_0\in[0.5,10]$, the laser wavelength $\lambda =0.8$ $\mu$m, the duration $\tau= 30$ fs and the laser spot size $w_0$
\begin{align}
    w_0\,[\mu\text{m}]= 38.2 \, \frac{\lambda\,[\mu \text{m}]}{a_0} \sqrt{\frac{\mathcal{E}_0\,\text{[J]}}{\tau\,\text{[30 fs]}}}.
    \label{eq:spotsize}
\end{align}
We simulate the interaction in 3D using the Locally Monochromatic
Approximation (LMA), in contrast to the previous sets of simulations which use the LCFA, because it is more accurate for low $a_0$~\cite{Heinzl_2020,10.1063/5.0159963}.
We see in \cref{fig:LUXEsweep}a) that if we select regions b-d) that the estimated largest allowed fluctuation in $a_0$ is between $2.6\%$ and $3.7\%$ for $5.0 < a_0 < 10.0$.
For reference, the LUXE experiment aims to achieve a precision of better than 5\% for the absolute intensity (1\% for the relative intensity) of the laser~\cite{Abramowicz2021}.

\subsection{\label{sec:3dsim} Collision stability tolerance}
In the previous section we saw that a lower positron yield, which is the predicted outcome of polarisation dependence, can be caused by fluctuations in $a_0$. However, a lower positron yield may also be caused by a misalignment of the colliding beams as the intersection area itself becomes smaller, which can interpreted effectively as a smaller $a_0$. 
In this section we will explore the effects of a spatial or a temporal offset between the laser and electron beam. We first consider the case where the two beams in a head-on collision are spatially misaligned by a transverse offset. 

\begin{figure}

    \includegraphics[width=1\linewidth]{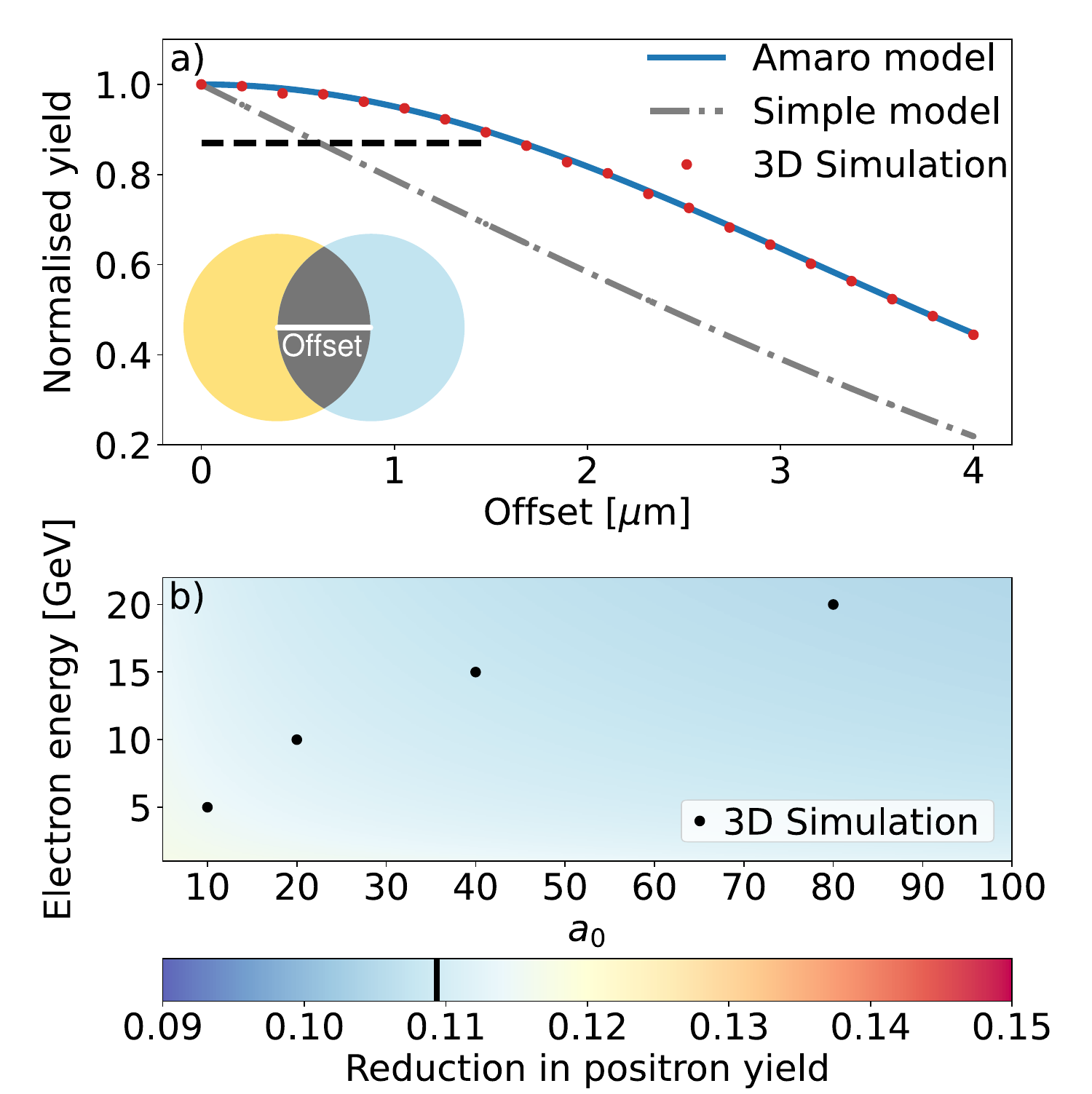}
    \caption{a) The positron yield, normalised to its value at zero offset, when an electron beam and a laser collide with given transverse offset: 3D simulation results (red points), analytical result from the \citet{Amaro_2021} model (blue line), and the normalised intersection area of the simple model (grey, dot-dashed line).
    The horizontal black dashed line indicates where the yield is reduced by $12\%$.
    b) The reduction in the yield predicted by the Amaro model at a 1.5 $\mu$m offset. The black dots indicate the parameters where 3D simulations are performed.}
    \label{fig:simple_intesection}
\end{figure}

We can model the reduction in the intersection area in a simple way, where the beam intensities are identical uniform disks with an radius of 3 $\mu$m, separated by an offset. The simple model is shown in \cref{fig:simple_intesection}a) where the yellow and light blue disks represent the laser and electron beams, and the grey area is the intersection of the two.

The normalised yield of pair production (normalised to zero offset) for this model is shown as a dashed-dotted line. Where we can see that the drop is linear and there is a 20\% reduction in yield after 1 $\mu$m, suggesting that the beam alignment has to be better than micron scale.

We compare this to 3D simulations of the collision between an electron beam of energy 10 GeV, bunch radius of 3 $\mu$m and a bunch length of 1 $\mu$m, and a laser with $a_0 =30$, a waist of 3 $\mu$m, a duration of 30 fs and a wavelength of 0.8 $\mu$m, for a given offset. The results of the simulations are shown in \cref{fig:simple_intesection}a) as red points. The reduction is not linear because the simple model neglects the complex intersection geometry of counter-propagating beams with Gaussian profiles, which simply is not well represented by uniform disks. 

However, we can also use the analytical models derived in \citet{Amaro_2021} to calculate the positron yield for different electron beam geometries, and with an spatial and temporal offset (transverse and longitudinal offset). Assuming that the electron bunch is short, and using the parameters above we integrate \cref{eq:total_prob} over the distribution of effective $a_0$ to get the yield, which we then normalise to the aligned setup.~\footnote{It should be noted that \cite{Amaro_2021} defines the electron beam radius differently from Ptarmigan which uses the RMS radius such that $R_\text{Ptarmigan} = \sqrt{2}R_\text{Amaro}$.} The results are shown in \cref{fig:simple_intesection}a) by the solid blue line and the analytical predictions are in good agreement with the 3D simulations for the different offsets. The offset that results in the yield dropping by more than 12\% is approximately 1.5 $\mu$m, which is represented by the black dashed line, which leaves little room for misalignment. This allowed offset changes very slowly with respect to $\mathcal{E}$ and $a_0$, as shown in \cref{fig:simple_intesection}b) where we have used the Amaro model to compute the reduction in the yield.
The results of 3D simulations for selected points (black dots) differ from the Amaro model by less then 2\%. We conclude that the misalignment needs to be less than half the beam radii in order to see the polarisation dependence in the positron yield.

\begin{figure}
    \centering
    \includegraphics[width=1\linewidth]{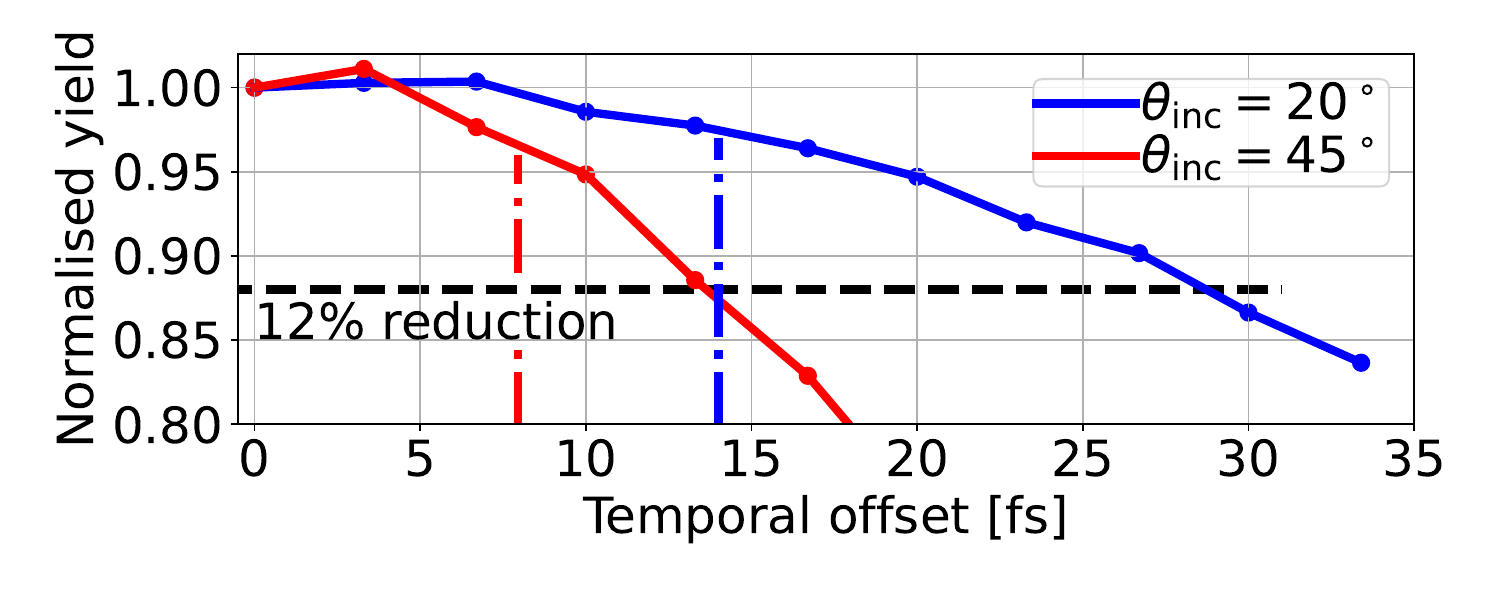}
    \caption{The reduction in the positron yield due to a temporal offset and at an incident angle $\theta_\text{inc}$ of the electron and laser beam. Blue points correspond to an incident angle of $20^\circ$ and red points to $45^\circ$. The 12\% yield reduction in order to distinguish the sought after polarization effects is marked with a dashed line and the vertical lines indicate the offset as predicted by \cref{eq:delta} where $a_0$ reduces by 5\%.}
    \label{fig:temporal}
\end{figure}

If the beams collide at an angle, the timing of the collision becomes important, because the electron bunch can miss the laser pulse, which it would not do in a head-on collision. Therefore, we consider the effect of a temporal misalignment between the electron and laser beam if they collide at an incident angle $\theta_\text{inc}$. We may analytically compute the effective $a_0$ in a collision at a incident angle $\theta_\text{inc}$ with a timing offset $\delta$. The effective laser intensity parameter $a$ as a function of space and time may be written as,
\begin{align}
    a &= a_0\exp{\left(-\frac{r^2}{w_0^2}\right)}\exp{\left[-\frac{\ln{2}\phi
    ^2}{2\pi^2n^2}\right]}\label{eq:a}.
\end{align}
Here $r^2= x^2 +y^2$, $w_0$ is the laser spot size, $\phi = 2\pi(ct - z)/\lambda$ is the phase, $\lambda$ is the laser wavelength and $n = c\tau/\lambda$ is the number of wavelengths in a FWHM pulse duration $\tau$.~Then $x = (\delta-ct)\sin{\theta_\text{inc}}$ and $z = (\delta-ct)\cos\theta_\text{inc}$. By maximising \cref{eq:a} with respect to time, we find the effective $a_0$,
\begin{align}
    a_{\text{max}} &= a_0 \exp{\left[-\frac{\delta^2\sin^2{\theta_\text{inc}}\ln{4}}{c^2\tau^2 \sin^2{\theta_\text{inc}} + w_0 ^2(1+\cos{\theta_\text{inc}})^2\ln4}\right]}.
\end{align}

The $\delta$ that gives rise to a change in the effective $a_0$ of $\Delta a_0$, assuming $\delta$ is small, is
\begin{align}
    \delta &= \sqrt{\frac{\Delta a_0\left(c^2\tau^2 \sin^2{\theta_\text{inc}} + w_0^2 (1+\cos\theta_\text{inc})^2\ln4\right)}{c^2\sin^2{\theta_\text{inc}}\ln{4}}}\label{eq:delta}.
\end{align}
To get a laser intensity fluctuation $\Delta a_0 \simeq 5\%$, given the parameters $w_0 = 3$ $\mu$m and $\tau =30$ fs, requires a temporal offset of 8 fs or 14 fs at the incident angles 20$^\circ$ and 45$^\circ$ respectively.

We verify this using simulations, where the collision parameters are the same as in \cref{fig:simple_intesection}, and in particular, the electron bunch size is $3$ $\mu$m. The temporal offset is translated to a spatial offset along the propagation axis for the electron bunch. \Cref{fig:temporal} shows the yield normalised to zero offset for the two incident angles $20^\circ$ (blue) and $45^\circ$ (red). A reduction in the normalised yield of 12\% is illustrated by the dashed line, below which the measurement of polarization dependence should be impossible. This occurs at an offset of 30 fs for an incident angle 20$^\circ$ and 15 fs at 45$^\circ$. These are larger than what \cref{eq:delta} predicts, but the ratios are consistent in that a larger $\theta_\text{inc}$ demands a better timing accuracy. The larger timing offset in simulation may be explained by the fact that \cref{eq:delta} assumes the electron beam size to be zero, which agrees with simulations, when the electron beam has a zero radius. We conclude that at a typical angle of 20$^\circ$, that the temporal offset should be less than half of the laser pulse duration.

\section{\label{sec:conclusion}Conclusion \text{ }}
Here we have re-considered the common assumption that emitted photons from Compton scattering (using a linearly polarised laser) can be treated as unpolarised. However, when we track the polarisation we find that 78\% of the emitted photons are polarised along the laser electric field (E-polarised). Since nonlinear Breit-Wheeler has a polarisation dependence which is in favour of B-polarised photons, we see an overestimation of the total number of positrons (on average 12\%) when treating photons as unpolarised. In principle this difference should be measurable, but we have shown that the laser intensity has to be known to better than a few per cent, and that the laser and electron beams cannot be misaligned by more than 1.5 $\mu$m, in order to be certain the reduction comes from polarisation dependence. This is experimentally challenging \cite{samarin}, but may be possible in future experiments that focus on beam stability. 

However, if the processes of Compton scattering and nonlinear Breit-Wheeler can be split up into different stages, such that the photon beam polarisation can be controlled with respect to the second laser beam, it would be easier to measure the polarisation dependence because the difference in yield is larger. The expected yield increase is 70\% when the polarisation is rotated 90$^\circ$ (from $E$- to $B$-polarisation)~\cite{2ndpaper}, which should be large enough to distinguish polarisation effects, even if the laser amplitude fluctuates by as much as 15\% (according to \cref{eq:a0fluct}).
A two-stage setup is more complicated, but is planned to be realised in the LUXE experiment \cite{Abramowicz2021}.
 
\section{Acknowledgement}
Simulations were performed on resources provided by the National Academic Infrastructure for Supercomputing in Sweden (NAISS), partially funded by the Swedish Research Council through grant agreement no. 2022-06725.

\bibliography{references.bib}

\end{document}